\documentclass{ws-procs9x6}

\begin{document}

\title{Non-Linear Effects in High Energy Hadronic 
Interactions}

\author{S.~Ostapchenko\footnote{\uppercase{N}ow at 
\uppercase{F}orschungszentrum \uppercase{K}arlsruhe, 
\uppercase{I}nstitut f\"ur \uppercase{K}ernphysik,
76021 \uppercase{K}arlsruhe,  \uppercase{G}ermany.}}

\address{Institut f\"ur Experimentelle Kernphysik, \\
         University of Karlsruhe,\\ 76021 Karlsruhe,  Germany \\
E-mail: serguei@ik.fzk.de}

\address{D.V. Skobeltsyn Institute of Nuclear Physics, \\
         Moscow State University,\\ 119992 Moscow, Russia}

\maketitle

\abstracts{
The influence of non-linear interaction effects on the characteristics of hadronic 
collisions and on the development of extensive air showers (EAS) is investigated.
Hadronic interactions are treated phenomenologically in the framework of Gribov's
Reggeon approach with non-linear corrections being described by enhanced
(Pomeron-Pomeron interaction) diagrams. A re-summation algorithm for higher order 
enhanced graphs is proposed. The approach is applied to develop a new
hadronic interaction model QGSJET-II,
treating non-linear effects explicitely  
 in individual hadronic and nuclear collisions. The model is applied to EAS
modelization and the obtained results are compared to the original QGSJET
 predictions. Possible consequences for EAS data interpretation
are discussed.
}

\section{Introduction\label{intro.sec} }

Currently the physics of hadronic interactions remains one of the most 
intriguing fields both for experimental and for theoretical research. Being the subject
of investigations in collider experiments, it has a different role in high energy
cosmic ray (CR) studies. There, a proper understanding of hadronic interaction
mechanisms is of vital importance for a correct interpretation of CR data,
 for a
reconstruction of energy spectra and of particle composition of the primary cosmic
radiation, and finally, for inferring informations on the sources of CR particles,
whose energies extend to ZeV region, exceeding by far those attainable at 
 man-made machines.

Despite a significant progress in QCD over the past two decades, only phenomenological
treatment is generally possible for minimum-bias
hadron-hadron (hadron-nucleus, nucleus-nucleus) collisions. 
In particular, Gribov's
Reggeon scheme\cite{gri68} proved to be a suitable framework for developing
successful  model approaches, 
like Quark-Gluon String or Dual Parton
models,\cite{kai82} which in turn provided a basis for a number
 of Monte Carlo (MC) generators of hadronic and nuclear collisions,
 extensively used in accelerator or CR
fields, e.g., DPMJET,\cite{ran95} VENUS,\cite{wer93} or
 QGSJET.\cite{kal93,kal94}

Still, all the mentioned MC models are based on a linear picture,
the elastic scattering amplitude being defined by (quasi-)eikonal contributions of 
independent Pomeron exchanges. Meanwhile, in the limit of very high energies one expects
a significant increase of parton density in colliding hadrons (nuclei), which gives rise
to essential non-linear interaction effects.\cite{glr} On the other hand, in Gribov's 
scheme the latter are traditionally described as Pomeron-Pomeron 
interactions.\cite{kan73,car74} Here we propose a re-summation procedure
both for enhanced contributions to the elastic scattering amplitude and for various
unitarity cuts of the corresponding diagrams. This gives us
 a possibility to account for
non-linear corrections when calculating hadron-hadron cross sections and to develop
a new MC model, QGSJET-II,\cite{pylos1} which treats non-linear effects
explicitely in individual hadronic and nuclear collisions.
The model is applied to simulate extensive air showers (EAS), induced by CR particles
in the atmosphere, which allows to investigate the influence of  
non-linear corrections on the air shower characteristics
 and to draw possible consequences for EAS data interpretation.

\section{Linear Scheme\label{lin-scheme}}

\begin{figure}[ht]
\begin{center}
\includegraphics[
  width=7cm,
  height=3cm]{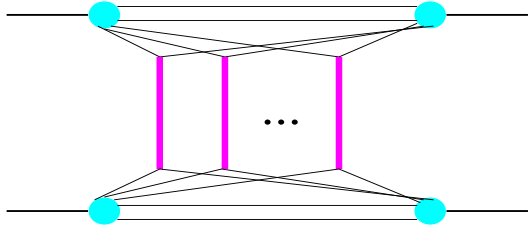}
  \end{center}
\caption{A general multi-Pomeron contribution to hadron-hadron scattering
amplitude. Elementary scattering processes (vertical thick lines)
are described as Pomeron exchanges; thin lines correspond to
constituent partons, to which Pomerons are coupled.\label{multiple}}
\end{figure}

Gribov's Reggeon approach\cite{gri68} describes a high energy
hadron-hadron collision as a multiple scattering process, 
where elementary re-scatterings are treated phenomenologically
as Pomeron exchanges -- Fig.~\ref{multiple}.%
Thus, hadron $a$ -- hadron $d$ elastic amplitude can be obtained summing
over multi-Pomeron exchange contributions\footnote{In the high energy
 limit all amplitudes
can be considered as pure imaginary.}\cite{kai82,bak76} (see also 
Ref. \refcite{dre01}):
\begin{eqnarray}
f_{ad}(s,b) &=& i\sum_{j,k} C_j^a\, C_k^d\, \sum_{n=1}^{\infty} \frac{1}{n!}
\int \! \left[\prod_{l=1}^n dx_l^+ dx_l^-\, \lambda_j^a \,\lambda_k^d \,
 G_{ad}^{\mathbb P}(x_l^+ x_l^- s,b)\right] \nonumber \\ 
&& \times N_a^{(n)}(x_1^+,...,x_n^+)\;N_d^{(n)}(x_1^-,...,x_n^-), \label{ampl}
\end{eqnarray}
where $s$ and $b$ are the c.m.~energy squared and the impact parameter
of the interaction, $G_{ad}^{\mathbb P}(x^+ x^- s,b)$ is the unintegrated
Pomeron exchange eikonal (for fixed values of Pomeron light cone 
momentum shares $x^{\pm}$), and $N_a^{(n)}(x_1,...,x_n)$ is the light cone
momentum distribution of constituent partons -- ``Pomeron ends'' (here,
quark-antiquark pairs). $C_j^a$, $\lambda_j^a$ are the relative weights
and strengths of diffractive eigenstates of hadron  $a$ in the Good-Walker
formalism;\cite{goo60} $\sum_j C_j^a=1$, $\sum_j C_j^a\,\lambda_j^a=1$. 
In particular, a two-component model ($j=1,2$) with one
passive component, $\lambda_2^a\equiv 0$, corresponds to the quasi-eikonal 
approach, where $\lambda_1^a\equiv 1/C_1^a$ is the shower enhancement 
coefficient.\cite{kai82}

Eq.~(\ref{ampl}) can be greatly simplified assuming that the integral over
light cone momenta can be factorized:\footnote{Here we neglect
energy-momentum correlations between multiple re-scatterings.\cite{hla01}}
\begin{eqnarray}
&& \int \left[\prod_{l=1}^n dx_l^+ dx_l^-\;
 G_{ad}^{\mathbb P}(x_l^+ x_l^- s,b)\right] \,
 N_a^{(n)}(x_1^+,...,x_n^+)\;N_d^{(n)}(x_1^-,...,x_n^-) \nonumber \\ 
&& = \left[\int dx^+ dx^-\,
 G_{ad}^{\mathbb P}(x^+ x^- s,b)\,N_a^{(1)}(x^+)\;N_d^{(1)}(x^-)\right]^n 
 \label{fact}
\end{eqnarray}

This leads to traditional eikonal formulas:
\begin{eqnarray}
f_{ad}(s,b) &=& i\sum_{j,k} C_j^a\, C_k^d\, 
\left[1-e^{- \lambda_j^a \,\lambda_k^d \, \chi_{ad}^{\mathbb P}(s,b)} \right]
\label{ampl-eik}
\\ \chi_{ad}^{\mathbb P}(s,b) &\equiv& \int dx^+ dx^-\,
 G_{ad}^{\mathbb P}(x^+ x^- s,b)\;N_a^{(1)}(x^+)\;N_d^{(1)}(x^-)
\label{chi-eik}
\end{eqnarray}
  
\begin{figure}[ht]
\begin{center}
\includegraphics[
  width=8cm,
  height=3cm]{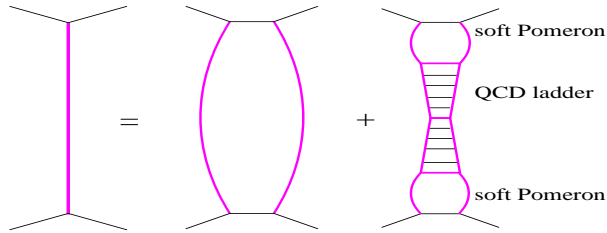}
  \end{center}
\caption{A ``general Pomeron'' (l.h.s.) consists of the ``soft'' and
the ``semi-hard'' Pomerons -- correspondingly the 1st and the 2nd
contributions on the r.h.s. \label{genpom} }
\end{figure}

In this scheme the Pomeron is an effective description of a microscopic
 parton cascade which mediates the interaction between the projectile
  and the target hadrons. It is convenient to divide the latter into two
  parts:``soft'' cascade of partons of small virtualities $|q^{2}|<Q_{0}^{2}$,
  and a perturbative parton evolution at  $|q^{2}|>Q_{0}^{2}$,
  where $Q_{0}^{2}$ is some cutoff for pQCD being applicable.
  Correspondingly, a ``general Pomeron'' will consist of two contributions:
``soft'' Pomeron for a pure non-perturbative process (all $|q^{2}|<Q_{0}^{2}$)
 and a ``semi-hard Pomeron'' for a cascade which at least partly develops
  in the high virtuality region 
  (some $|q^{2}|>Q_{0}^{2}$)\cite{kal94,dre01,dre99}
  -- Fig. \ref{genpom}:
   \begin{equation}
G_{ad}^{\mathbb P}(\hat s,b)=G_{ad}^{{\mathbb P}_{\rm soft}}(\hat s,b)
+G_{ad}^{{\mathbb P}_{\rm sh}}(\hat s,b)\label{chi-tot}
\end{equation}
The advantage of such a procedure is that in very high energy limit the ``general Pomeron''
is dominated by its semi-hard component and the energy dependence of the eikonal
$G_{ad}^{\mathbb P}(\hat s,b)$ is governed asymptotically 
by the perturbative QCD evolution.

The soft Pomeron eikonal can chosen in the usual form:\cite{kai82}
\begin{equation}
G_{ad}^{{\mathbb P}_{\rm soft}}(\hat s,b)
=\frac{\gamma_{a}\,\gamma_{d}\,(\hat s/s_{0})^{\alpha_{{\mathbb P}}(0)-1}}
{R_{a}^{2}+R_{d}^{2}+\alpha_{{\mathbb P}}'(0)\,\ln \!\frac{\hat s}{s_{0}}}
e^{-\frac{b^{2}}
 {4\left(R_{a}^{2}+R_{d}^{2}+\alpha_{{\mathbb P}}'(0)\,
 \ln \!\frac{\hat s}{s_{0}}\right)}}, \label{G-soft}
 \end{equation}
where  $s_{0}\simeq 1$ GeV$^{2}$ is the hadronic mass scale, $\alpha_{{\mathbb P}}(0)$
and $\alpha_{{\mathbb P}}'(0)$ are the intercept and the slope of the
Pomeron Regge trajectory, and $\gamma_{a}$, $R_{a}^{2}$ are the
coupling and the slope of Pomeron-hadron $a$ interaction vertex.

The dominant contribution to the semi-hard Pomeron comes from hard scattering
of gluons and sea quarks\footnote{For brevity, hard scattering of 
valence quarks is not discussed explicitely.}
 and can be represented by a piece of QCD ladder
sandwiched between two soft Pomerons\footnote{In general, 
one may consider an arbitrary number of $t$-channel iterations
of soft and hard Pomerons.\cite{tan94}} -- Fig. \ref{genpom}; 
the corresponding eikonal may be written as\cite{kal94,dre01,dre99}
\begin{eqnarray}
G_{ad}^{{\mathbb P}_{\rm sh}}(\hat s,b) &=& \frac{1}{2}
\sum_{I,J=g,\bar q} \int \!d^2b' \int \!\frac{dx_h^+}{x_h^+}\frac{dx_h^-}{dx_h^-}
\; \sigma _{IJ}^{\rm QCD}(x_h^+ x_h^- \hat s,Q_0^2) \nonumber \\
& & \times G_{aI}^{{\mathbb P}_{\rm soft}}(s_0/x_h^+,b') \;
G_{dJ}^{{\mathbb P}_{\rm soft}}(s_0/x_h^-,|\vec b - \vec b'|) 
\label{G-sh}
\end{eqnarray}
Here $\sigma _{IJ}^{\rm QCD}(x_h^+ x_h^- \hat s,Q_0^2)$ is the contribution of 
the parton ladder with the virtuality cutoff $|q^2|>Q_0^2$;
 $I,J$ and $x_h^+,x_h^-$ are the types and the
 relative light cone momentum shares of the ladder
leg partons. The eikonal $G_{aI}^{{\mathbb P}_{\rm soft}}(\hat s,b')$, 
describing parton $I$ momentum and impact parameter distribution in the soft Pomeron
at virtuality scale $Q_0^2$, is defined by Eq.~(\ref{G-soft}),
neglecting the small slope of Pomeron-parton coupling 
$R_I^2\sim 1/Q_0^2\sim 0$ and using a parameterized Pomeron-parton $I$ 
  vertex $\gamma _I(x_h)$, $x_h=s_0/\hat s$. Usual hadronic parton momentum
distribution functions (PDFs) can be obtained convoluting 
$G_{aI}^{{\mathbb P}_{\rm soft}}$ with the constituent parton distribution $N_a^{(1)}$
and integrating over the parton impact parameter $b'$:
\begin{equation}
x\, f_{I/a}(x,Q_{0}^{2})=\int \!d^2b' \int _x^1 \!dx^+\; N_a^{(1)}(x^+) \;
G_{aI}^{{\mathbb P}_{\rm soft}}(s_0\,x^+/x,b')\label{pdf-q0}
\end{equation}

Thus, the contribution of semi-hard processes to the integrated eikonal (\ref{chi-eik}),
integrated over the impact parameter $b$, can be written\cite{dre01,dre99}
 as a convolution of hadronic PDFs $f_{I/a(J/d)}(x,Q^{2})$ (obtained by evolving
the input PDFs (\ref{pdf-q0}) from $Q_0^2$ to $Q^2$) with the parton
 scatter cross section $d\sigma_{ij}^{2\rightarrow2}/dp_{t}^{2}$:
\begin{eqnarray}
\int \!d^2b\,\chi_{ad}^{{\mathbb P}_{\rm sh}}(s,b) &=& \frac 1 2
\int\! dx_h^{+}\,dx_h^{-}\int\!dp_{t}^{2} 
\left\{ K \right. \sum_{I,J}
\frac{d\sigma_{ij}^{2\rightarrow2}(x_h^{+}x_h^{-}s,p_{t}^{2})}{dp_{t}^{2}}
 \nonumber \\ && \left.
\times f_{I/a}(x_h^{+},M_{F}^{2})\: f_{J/d}(x_h^{-},M_{F}^{2})\:
\Theta(M_{F}^{2}-Q_{0}^{2}) \right\},\label{chi-sh-int}
\end{eqnarray}
with $p_{t}$ being parton transverse momentum in the hard process, 
$M_{F}^{2}$ -- the factorization scale (here $M_{F}^{2}=p_{t}^{2}/4$),
and with the factor $K$ accounting for higher order corrections.
Clearly, the integrand in the curly brackets on the r.h.s.~of 
Eq.~(\ref{chi-sh-int}) defines inclusive jet production cross section.

Knowing the elastic scattering amplitude, Eq.~(\ref{ampl-eik}), 
one can calculate total and elastic cross sections, 
elastic scattering slope for the interaction, etc.
Moreover, cutting the corresponding diagrams of Fig.~\ref{multiple} according
to the  Abramovskii-Gribov-Kancheli (AGK) cutting rules\cite{agk} and collecting
contributions of cuts of certain topologies, one can obtain
 relative weights for various configurations of the interaction.\cite{kai82}
The latter gives a possibility to develop MC generation procedures for 
hadron-hadron,
hadron-nucleus, and nucleus-nucleus collisions.\cite{kal93,kal94,dre99}

\section{Pomeron-Pomeron Interactions}

\begin{figure}[ht]
\begin{center}
\includegraphics[
  width=8cm,
  height=3cm]{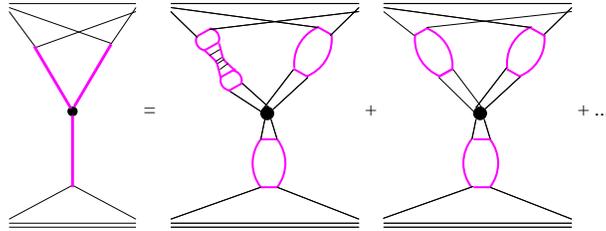}
  \end{center}
\caption{Contributions to the triple-Pomeron vertex from interactions between
soft and semi-hard Pomerons.\label{3p-vertex}}
\end{figure}

The above-described linear picture cease to be valid in
the ``dense'' regime, i.e.~in the limit of high energies
and small impact parameters of the interaction. There, a large number
of elementary scattering processes occurs and corresponding underlying
parton cascades largely overlap and interact with each other. Such
effects are traditionally described by enhanced 
diagrams.\cite{kan73,car74} To develop a dynamic scheme it is assumed
that Pomeron-Pomeron interactions are dominated by partonic
processes at comparatively low virtualities,\cite{pylos1,dre01}
 $|q^{2}|<Q_{0}^{2}$,
 and  that Pomeron-Pomeron coupling at larger virtualities
becomes important only after reaching the saturation regime for parton densities
at the scale $Q_0^2$.
Thus, we develop a scheme where multi-Pomeron vertexes involve only interactions
 between soft Pomerons or between ``soft ends'' of semi-hard Pomerons --
Fig.~\ref{3p-vertex}.

\begin{figure}[ht]
\begin{center}
\includegraphics[
  width=3.5cm,
  height=2.5cm]{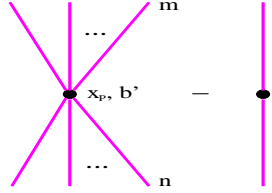}
  \end{center}
\caption{Lowest order enhanced graphs; constituent partons not shown
explicitely.\label{1-3P}}
\end{figure}

Basically we shall stay close to the $\pi$-meson dominance picture,\cite{car74}
where all the vertexes $g_{mn}$ for the transition of $m$ into $n$ Pomerons 
have been expressed via a single additional constant $r_{3{\mathbb P}}$
 and where an asymptotic re-summation procedure has been proposed.\cite{kai86}
In particular, for the lowest in $r_{3{\mathbb P}}$ contribution with
 only one multi-Pomeron vertex (Fig.~\ref{1-3P})
one can obtain, summing over $m,n\geq1$ and subtracting the term
with $m=n=1$ (Pomeron self-coupling):\cite{kai86}
\begin{eqnarray}
\Delta\chi_{ad}^{(1)(j,k)}(s,b) &= & 
\frac{r_{3{\mathbb P}}}{\lambda_j^a \,\lambda_k^d\,\gamma_{\mathbb P}^3} 
\int_{\frac{s_{0}}{s}}^{1}\!\frac{dx_{{\mathbb P}}}{x_{{\mathbb P}}}
\int\! d^{2}b' \,\left\{ \left[1-e^{-\lambda_j^a \,
\chi_{a{\mathbb P}}^{\mathbb P}(s_{0}/x_{{\mathbb P}},b') }\right]
 \right.\nonumber \\
&& \times \left[1-e^{-\lambda_k^d \,
\chi_{d{\mathbb P}}^{\mathbb P}(x_{{\mathbb P}}s,|\vec{b}-\vec{b}'|) }
\right]\nonumber \\
&& \left.-\lambda_j^a \,\lambda_k^d \;
\chi_{a{\mathbb P}}^{\mathbb P}(s_{0}/x_{{\mathbb P}},b')\;
\chi_{d{\mathbb P}}^{\mathbb P}(x_{{\mathbb P}}s,|\vec{b}-\vec{b}'|)\right\},
\label{3p-(1)}
\end{eqnarray}
with $\chi_{a{\mathbb P}}^{\mathbb P}(\hat{s},b')\equiv
\chi_{a\pi}^{\mathbb P}(\hat{s},b')$, 
$\gamma _{\mathbb P}=\gamma_{\pi}$; here and below the indexes $j,k$ refer
to the diffractive eigenstates of hadrons $a,d$ correspondingly.
It is easy to see that in the large $s$, small $b$ limit 
$\Delta\chi_{ad}^{(1)(j,k)}(s,b)$ is dominated by the last term 
in the integrand, i.e.~by the subtracted self-coupling contribution.
Therefore, asymptotically it was sufficient to consider a small 
sub-set of enhanced graphs, which can be obtained from the
 one in Fig.~\ref{1-3P} iterating   multi-Pomeron vertex
in  $t$-channel. In the ``dense'' limit re-summation of those
 diagrams reduces to the sum over subtracted Pomeron self-couplings 
 and, after neglecting the slope of the
multi-Pomeron vertex, $R_{\mathbb P}^2\sim 0$, results in a re-normalization
of the Pomeron intercept,\cite{kai86}
$\tilde \alpha_{{\mathbb P}}(0)=
\alpha_{{\mathbb P}}(0)-4\pi\,r_{3{\mathbb P}}/\gamma_{\mathbb P}$.

\begin{figure}[ht]
\begin{center}
\includegraphics[
  width=5cm,
  height=3cm]{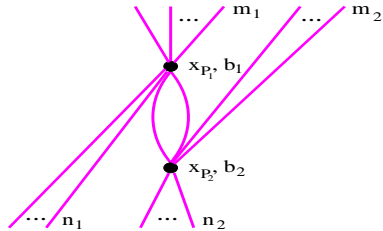}
  \end{center}
\caption{An example of a ``loop'' graph.\label{loop}}
\end{figure}

In this work we also assume an eikonal structure of the multi-Pomeron vertexes,
$\gamma _{{\mathbb P}-{\mathbb P}}^{(m,n)}
=r_{3{\mathbb P}}\,\gamma_{\mathbb P}^{m+n-3}$, 
but treat $\gamma_{\mathbb P}$ as a free parameter and neglect the vertex slope,
 $R_{\mathbb P}^2\sim 0$. On the other hand, we neglect a momentum
  spread of ``Pomeron ends'' in the vertexes and define the eikonal 
  $\chi_{a{\mathbb P}}^{\mathbb P}$  as
\begin{equation}
\chi_{a{\mathbb P}}^{\mathbb P}(\hat{s},b') = \int _{s_0/\hat s}^1 
\!dx^+ \;N_a^{(1)}(x^+) \;
G_{a{\mathbb P}}^{\mathbb P}(x^+\hat s,b'),
\end{equation}
where $G_{a{\mathbb P}}^{\mathbb P}$ is given by 
Eqs.~(\ref{chi-tot}-\ref{G-sh}), with $d\rightarrow \mathbb P$. 
Correspondingly, for a Pomeron exchanged between two internal vertexes,
e.g., in Fig.~\ref{loop},
 we just use $G_{{\mathbb P}{\mathbb P}}^{\mathbb P}(s_0\,
 x_{{\mathbb P}_k}/x_{{\mathbb P}_{k+1}},|\vec{b}_k-\vec{b}_{k+1}|)$,
 the latter being defined by Eqs.~(\ref{chi-tot}-\ref{G-sh})
  with $ad\rightarrow \mathbb P\mathbb P$.

\begin{figure}[ht]
\begin{center}
\includegraphics[
  width=4cm,
  height=3cm]{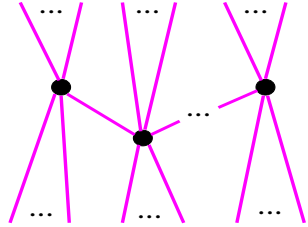}
  \end{center}
\caption{An example of a ``chess-board'' graph; only three vertexes
are shown explicitely.\label{chess}}
\end{figure}

Our goal is to  develop a re-summation procedure which  assures 
a smooth transition between
the ``dilute'' (small $s$, large $b$) and ``dense'' limits and  accounts for all
essential contributions in both cases. Thus, we shall only neglect 
``loop'' diagrams (Fig.~\ref{loop}) and ``chess-board'' graphs
with more than three vertexes in a horizontal row -- Fig.~\ref{chess}.
The neglected diagrams give small contribution
at low energies and/or large impact parameters, being proportional to large 
 powers ($\geq 2$) of $r_{3{\mathbb P}}$. On the other hand, 
 in the ``dense'' limit they are strongly 
suppressed by exponential factors.\cite{kai86} For the example of 
Fig.~\ref{loop}, summing over any number $n_1$ of Pomerons exchanged
 between the target and the upper
multi-Pomeron vertex (for the other vertex the procedure is similar), 
we can obtain a factor
\begin{equation}
\sum _{n_1=0}^{\infty}\frac{\left[-\lambda_k^d \,
\chi_{d{\mathbb P}}^{\mathbb P}(x_{{\mathbb P}_1}s,|\vec{b}-\vec{b}_1|) 
\right]^{n_1}}{n_1!}
=e^{-\lambda_k^d \,
\chi_{d{\mathbb P}}^{\mathbb P}(x_{{\mathbb P}_1}s,|\vec{b}-\vec{b}_1|) }
\end{equation}

We start with obtaining a ``fan'' diagram contribution 
$\chi_{a}^{{\rm fan}(j)}$ via 
a recursive equation -- Fig.~\ref{ffan}:
\begin{figure}[ht]
\begin{center}
\includegraphics[
  width=7cm,
  height=3cm]{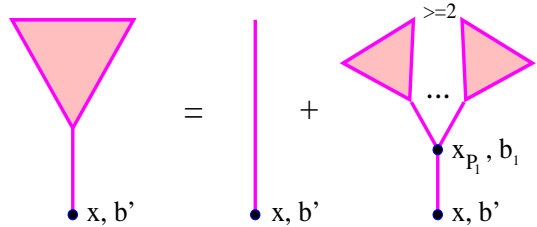}
  \end{center}
\caption{Recursive equation for a ``fan'' diagram contribution 
$\chi_{a}^{{\rm fan}(j)}(\hat{s},b')$,
$\hat{s}=s_{0}/x$.\label{ffan}}
\end{figure}
\begin{eqnarray}
\chi_{a}^{{\rm fan}(j)}\!(\hat{s},b')&= & 
\chi_{a{\mathbb P}}^{\mathbb P}(\hat{s},b')
-\frac{r_{3{\mathbb P}}}{\lambda_j^{a}\, \gamma_{\mathbb P}^3}
\int_{\frac{s_{0}}{\hat{s}}}^{1}\!\frac{dx_{{\mathbb P}_1}}{x_{{\mathbb P}_1}} 
\int\! d^{2}b_1\,G_{{\mathbb P}{\mathbb P}}^{\mathbb P}
(x_{{\mathbb P}_1}\hat{s},|\vec{b}'-\vec{b}_1|)\nonumber \\
&& \times \sum_{m=2}^{\infty}\,
\frac{\left[-\lambda_j^{a}\: \chi_{a}^{{\rm fan}(j)}
\!(s_{0}/x_{{\mathbb P}_1},b_1)
\right]^{m}}{m!} \nonumber \\
&=&  \chi_{a{\mathbb P}}^{\mathbb P}(\hat{s},b')
+\frac{r_{3{\mathbb P}}}{\lambda_j^{a}\, \gamma_{\mathbb P}^3}
\int_{\frac{s_{0}}{\hat{s}}}^{1}\!\frac{dx_{{\mathbb P}_1}}{x_{{\mathbb P}_1}} 
\int\! d^{2}b_1\;G_{{\mathbb P}{\mathbb P}}^{\mathbb P}
(x_{{\mathbb P}_1}\hat{s},|\vec{b}'-\vec{b}_1|)\nonumber \\
&& \times 
\left[1-e^{-\lambda_j^{a}\, \chi_{a}^{{\rm fan}(j)}
\!(s_{0}/x_{{\mathbb P}_1},b_1)}
-\lambda_j^{a}\, \chi_{a}^{{\rm fan}(j)}\!(s_{0}/x_{{\mathbb P}_1},b_1)\right]
\label{fan}
\end{eqnarray}

\begin{figure}[ht]
\begin{center}\includegraphics[%
  width=8cm,
  height=3cm]{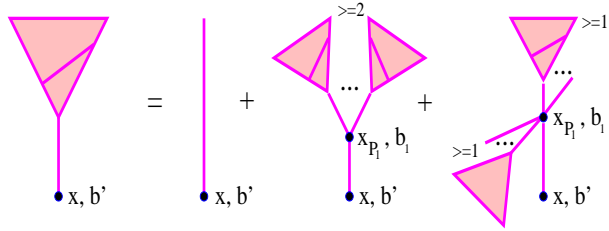}\end{center}
\caption{Recursive equation for a ``generalized fan''
 $\chi_{ad}^{{\rm G-fan}(j,k)}(\hat{s},\vec{b}',s,\vec{b})$,
$\hat{s}=s_{0}/x$.\label{freve}}
\end{figure}

Also we  introduce a ``generalized fan'' contribution $\chi_{ad}^{{\rm G-fan}(j,k)}$
via a recursive equation -- Fig.~\ref{freve}:
\begin{eqnarray}
\chi_{ad}^{{\rm G-fan}(j,k)}\!(\hat{s},\vec{b}',s,\vec{b}) &=&
 \chi_{a{\mathbb P}}^{\mathbb P}(\hat{s},b')
+\frac{r_{3{\mathbb P}}}{\lambda_j^{a}\, \gamma_{\mathbb P}^3}
\int_{\frac{s_{0}}{\hat{s}}}^{1}\!\frac{dx_{{\mathbb P}_1}}{x_{{\mathbb P}_1}}
\int\! d^{2}b_1 \nonumber \\
&& \times G_{{\mathbb P}{\mathbb P}}^{\mathbb P}
(x_{{\mathbb P}_1}\hat{s},|\vec{b}'-\vec{b}_1|) \,
\left\{ e^{-\lambda_k^{d}\,
\chi_{d}^{{\rm fan}(k)}\!(x_{{\mathbb P}_1}s,|\vec{b}-\vec{b}_1|)}
\right.\nonumber \\
&& \times
\left[1-e^{-\lambda_j^{a}\,
\chi_{ad}^{{\rm G-fan}(j,k)}\!
(s_{0}/x_{{\mathbb P}_1},\vec b_1,s,\vec{b})} \right] \nonumber \\
&& \left.
- \lambda_j^{a}\:
\chi_{ad}^{{\rm G-fan}(j,k)}\!
(s_{0}/x_{{\mathbb P}_1},\vec b_1,s,\vec{b})\right\}
 \label{g-fan}
\end{eqnarray}

\begin{figure}[htb]
\begin{center}\includegraphics[%
  width=7cm,
  height=2.5cm]{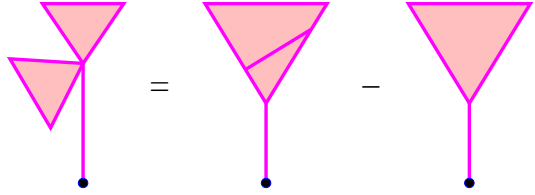}\end{center}
\caption{Symbolic notation for the difference between the ``generalized
fan'' and ``fan'' contributions, 
$\chi_{ad}^{{\rm G-fan}(j,k)}-\chi_{a}^{{\rm fan}(j)}$.\label{fdif}}
\end{figure}

The difference between $\chi_{ad}^{{\rm G-fan}(j,k)}$ and 
$\chi_{a}^{{\rm fan}(j)}$ (Fig.~\ref{fdif}) is due to vertexes 
with both ``fans'' connected to the projectile
and ones connected to the target. 

\begin{figure*}[ht]
\begin{center}\includegraphics[%
  width=11cm,
  height=2.7cm]{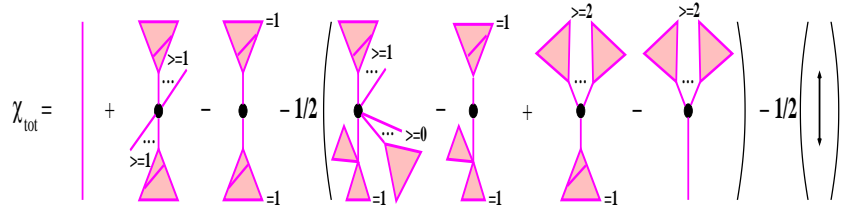}\end{center}
\caption{Total hadron-hadron eikonal, including enhanced diagram contributions.
\label{feiktot}}
\end{figure*}

Now we can obtain total eikonal for an ``elementary scattering'' process, 
including all essential enhanced
diagram contributions\cite{pylos1} -- Fig.~\ref{feiktot}:
\begin{eqnarray}
\chi_{ad}^{{\rm tot}(j,k)}\!(s,b) &=& \chi_{ad}^{\mathbb P}(s,b)
+\frac{r_{3{\mathbb P}}}{\lambda_j^{a}\, \lambda_k^{d}\, \gamma_{\mathbb P}^3}
\int_{\frac{s_{0}}{s}}^{1}\!\!\frac{dx_{{\mathbb P}}}{x_{{\mathbb P}}}
\int\!\! d^{2}b' \left\{ \left(1-e^{-\lambda_j^{a}\,
 \chi_{ad}^{{\rm G-fan}(j,k)}}\right) \right.\nonumber \\
&& \times \left(1-e^{-\lambda_k^{d}\, \chi_{da}^{{\rm G-fan}(k,j)}}\right)
- \lambda_j^{a}\, \lambda_k^{d}\: \chi_{ad}^{{\rm G-fan}(j,k)}\,
\chi_{da}^{{\rm G-fan}(k,j)}
\nonumber \\
&& -\left[\frac{\lambda_k^{d}}{2}
\left( \chi_{da}^{{\rm G-fan}(k,j)} -\chi_{d}^{{\rm fan}(k)}\right)\,
\left(\left(1-e^{-\lambda_j^{a}\,
\chi_{ad}^{{\rm G-fan}(j,k)}}\right)
\right. \right. \nonumber \\
&& \left. \times e^{-\lambda_k^{d}\,
 \chi_{d}^{{\rm fan}(k)}}
-\lambda_j^{a}\, 
\chi_{ad}^{{\rm G-fan}(j,k)} \right)
+\frac{\lambda_k^{d}}{2}
\left(\chi_{da}^{{\rm G-fan}(k,j)}
-\chi_{d{\mathbb P}}^{\mathbb P} \right) \nonumber \\
&& \left.\left.\times\left(1-e^{-\lambda_j^{a}\, 
\chi_{a}^{{\rm fan}(j)}} -\lambda_j^{a}\, 
\chi_{a}^{{\rm fan}(j)} \right)\right]
-\left[a,j\leftrightarrow d,k\right]\right\} \!,\label{chi-ad-tot}
\end{eqnarray}
where 
$\chi_{ad}^{{\rm G-fan}(j,k)}
=\chi_{ad}^{{\rm G-fan}(j,k)}(s_{0}/x_{{\mathbb P}},\vec{b}',s,\vec{b})$,
$\chi_{a}^{{\rm fan}(j)}=\chi_{a}^{{\rm fan}(j)}(s_{0}/x_{{\mathbb P}},b')$,
$\chi_{a{\mathbb P}}^{\mathbb P}
=\chi_{a{\mathbb P}}^{\mathbb P}(s_{0}/x_{{\mathbb P}},b')$,
$\chi_{da}^{{\rm G-fan}(k,j)}
=\chi_{da}^{{\rm G-fan}(k,j)}(x_{{\mathbb P}}\,s,\vec{b}-\vec{b}',s,\vec{b})$,
$\chi_{d}^{{\rm fan}(k)}=\chi_{d}^{{\rm fan}(k)}(x_{{\mathbb P}}\,s,|\vec{b}-\vec{b}'|)$,
$\chi_{d{\mathbb P}}^{\mathbb P}
=\chi_{d{\mathbb P}}^{\mathbb P}(x_{{\mathbb P}}\,s,|\vec{b}-\vec{b}'|)$.

The first term on the r.h.s.~in Fig.~\ref{feiktot} is a simple
Pomeron exchange ($\chi_{ad}^{\mathbb P}$); the second graph contains a multi-Pomeron
vertex which couples together any number ($\geq 1$) of ``generalized fans'' connected 
to the projectile hadron and any number ($\geq 1$) of those connected to the target;
the third graph subtracts Pomeron self-coupling contribution; the other terms
on the r.h.s.~in Fig.~\ref{feiktot} correct for double counts of the same diagrams
in the second graph. The latter can be verified using  
Figs.~\ref{ffan}--\ref{fdif}. 

Thus, hadron-hadron elastic scattering amplitude $f_{ad}(s,b)$ can be calculated 
using Eq.~(\ref{ampl-eik}), with the simple Pomeron eikonal
$\chi_{ad}^{\mathbb P}(s,b)$ being replaced by 
$\chi_{ad}^{\rm{tot}(j,k)}(s,b)$, corresponding to the full set of graphs of  
Fig.~\ref{feiktot}. To describe particle production 
one has to consider unitarity cuts of the amplitude
$f_{ad}(s,b)$. Applying AGK cutting rules,\cite{agk} one can obtain
 a set of all cuts of diagrams of Fig.~\ref{feiktot}, with the eikonal
  $\chi_{ad}^{\rm{tot-cut}(j,k)}(s,b)\simeq 
\chi_{ad}^{\rm{tot}(j,k)}(s,b)$,\footnote{Here the
equality between the uncut and cut eikonals is approximate due to
a somewhat different selection of neglected ``chess-board'' 
graphs in the two cases.\cite{pylos1}}
 in a similar form, i.e.~as a contribution of one cut Pomeron exchange, the
latter containing all relevant screening corrections 
(any number of multi-Pomeron vertexes along the same
 cut Pomeron line, each vertex having any number but at least one 
uncut ``fan'' connected to the projectile or to the target); plus a graph with
 a multi-Pomeron vertex, which couples together any number ($\geq 1$) of cut 
``generalized fans'' (including diffractive cuts) connected 
to the projectile hadron and any number ($\geq 1$) of those 
connected to the target, with all cut Pomeron lines being ``dressed'' with
screening corrections in a similar way;
minus double counting contributions.\cite{pylos1} 
This allows to obtain positively
defined probabilities for various configurations of the interaction
 and to generate the latter via a MC method: starting from sampling
  (at a given impact parameter $b$ and for given diffractive eigenstates $j,k$)
 a number of ``elementary'' particle production processes -- according to the eikonal
$\chi_{ad}^{\rm{tot-cut}(j,k)}(s,b)$, 
and choosing a particular sub-configuration for
the latter -- just one cut Pomeron exchange or a number of cut 
``generalized fans'' emerging from the same multi-Pomeron vertex;
 in the latter case for each cut ``fan''
one generates further multi-Pomeron vertexes in an iterative fashion,
 according to the corresponding probabilities, until the process is terminated. 
 Finally, like in the original linear scheme,\cite{kal93,kal94,dre99} 
 one performs energy-momentum sharing between constituent partons of 
 the projectile and of the target, to which cut Pomerons are connected, 
 and finishes with the treatment of particle
production from all cut Pomerons.
The described scheme easily generalizes to hadron-nucleus and nucleus-nucleus
collisions, without introducing any additional parameters.
There, screening effects are enhanced by $A$-factors, as different Pomerons
belonging to the same enhanced graph may couple to different nucleons of the
projectile (target).

\section{Cross Sections and Structure Functions}
 We can also calculate screening corrections to sea quark and gluon PDFs
at the initial virtuality scale $Q^2_0$, 
which come only from ``fan''-type diagrams (neglecting Pomeron ``loops'').
 Thus, parton momentum distributions,
being defined in the linear scheme by Eq.~(\ref{pdf-q0}), are now described
by the diagrams of Fig.~\ref{ffan}, with the corresponding contributions 
(Eq.~(\ref{fan})) being
summed over hadron $a$ diffractive eigenstates $j$ (with weights 
$C_j^a\,\lambda_j^{a}$),
 integrated over $b'$, and with the down-most
vertex being replaced by the Pomeron-parton coupling: 
\begin{eqnarray}
x\, f^{\rm tot}_{I/a}(x,Q_{0}^{2})& =&
\int \!\!d^2b' \left\{ \int _x^1 \!\!dx^+ \; N_a^{(1)}(x^+) \;
G_{aI}^{{\mathbb P}_{\rm soft}}(s_0\,x^+/x,b') \right. \nonumber \\
&& +\frac{r_{3{\mathbb P}}}{\gamma_{\mathbb P}^3}\sum _j C_j^{a}
\int_{x}^{1}\!\frac{dx_{{\mathbb P}_1}}{x_{{\mathbb P}_1}} 
\int\! d^{2}b_1\:
G_{{\mathbb P}I}^{{\mathbb P}_{\rm soft}}
(s_0\,x_{{\mathbb P}_1}/x,|\vec{b}'-\vec{b}_1|)\nonumber \\
&& \left. \times 
\left[1-e^{-\lambda_j^{a} 
\chi_{a}^{{\rm fan}(j)}\!(s_{0}/x_{{\mathbb P}_1},b_1)}
-\lambda_j^{a} \chi_{a}^{{\rm fan}(j)}\!(s_{0}/x_{{\mathbb P}_1},b_1)
\right]\right\}
\label{pdf-scr}
\end{eqnarray}

\begin{figure}[ht]
\begin{center}\includegraphics[%
  width=11cm,
  height=3.5cm]{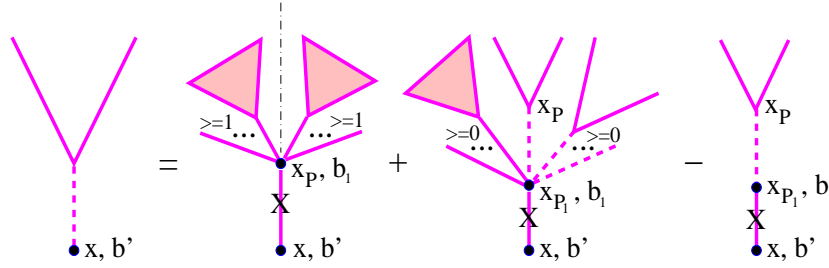}\end{center}
\caption{Recursive equation for the diffractive cut 
$\chi_{a}^{{\rm DD}(j)}\!(\hat s,x_{{\mathbb P}},b')$ of ``fan'' diagrams,
$\hat{s}=s_{0}/x$. Cut Pomerons are marked by crosses.\label{fan-dd}}
\end{figure}

Similarly one can express diffractive PDFs $f^{\rm DD}_{I/a}(x,x_{{\mathbb P}},Q_{0}^{2})$ 
for a large rapidity gap process in deep inelastic scattering
(with $y_{\rm gap}=-\ln x_{{\mathbb P}}$)
 via the contribution of diffractive cuts of ``fan'' diagrams 
 $\chi_{a}^{{\rm DD}(j)}(\hat s,x_{{\mathbb P}},b')$. The latter is
 obtained cutting the diagrams of Fig.~\ref{ffan} according to the AGK
 rules and keeping contributions
 of cuts with the given rapidity gap, which yields again a recursive 
 equation -- Fig.~\ref{fan-dd}:
\begin{eqnarray}
\chi_{a}^{{\rm DD}(j)}\!(\hat s,x_{{\mathbb P}},b')& =&
  \frac{r_{3{\mathbb P}}}{\lambda_j^{a}\, \gamma_{\mathbb P}^3} 
\int \!d^2b_1 \left\{
\frac{1}{2} \left[1-e^{-\lambda_j^{a}\, 
\chi_{a}^{{\rm fan}(j)}\!(s_{0}/x_{{\mathbb P}},b_1)}\right]^2 \right. \nonumber \\
&& \times 
G_{{\mathbb P}{\mathbb P}}^{\mathbb P}(x_{{\mathbb P}}\, \hat s,
|\vec{b}'-\vec{b}_1|)  
+ \int_{x}^{x_{{\mathbb P}}}\!  \frac{dx_{{\mathbb P}_1}}{x_{{\mathbb P}_1}} \,
G_{{\mathbb P}{\mathbb P}}^{\mathbb P} (x_{{\mathbb P}_1} \hat s,
|\vec{b}'-\vec{b}_1|)
 \nonumber \\ && \times 
\lambda_j^{a}\, 
\chi_{a}^{{\rm DD}(j)}\!\left(\frac{s_{0}}{x_{{\mathbb P}_1}},
x_{{\mathbb P}},b_1\right)
 \,\left[\exp \left(-\lambda_j^{a}\, 
\chi_{a}^{{\rm fan}(j)}\!\left(\frac{s_{0}}{x_{{\mathbb P}_1}},b_1\right)
  \right.\right. \nonumber \\
&&  +\int_{x_{{\mathbb P}_1}}^{x_{{\mathbb P}}}\! 
 \frac{dx_{{\mathbb P}_2}}{x_{{\mathbb P}_2}} \!
 \left. \left. \left.\lambda_j^{a}\, \chi_{a}^{{\rm DD}(j)}
\!\left(\frac{s_{0}}{x_{{\mathbb P}_1}},x_{{\mathbb P}_2},b_1\right)
\right)-1 \right] \right\}
\label{fan-difr}
\end{eqnarray}
Correspondingly, the diffractive PDFs 
$f^{\rm DD}_{I/a}(x,x_{{\mathbb P}},Q_{0}^{2})$ are described
by the diagrams of Fig.~\ref{fan-dd}, with the corresponding contributions
(Eq. (\ref{fan-difr})) being
summed over hadron $a$ diffractive eigenstates, integrated over $b'$, and 
with the down-most vertex being replaced by the Pomeron-parton $I$ coupling,
i.e.~with the eikonal $G_{{\mathbb P}{\mathbb P}}^{\mathbb P}$ being replaced by
$G_{{\mathbb P}I}^{{\mathbb P}_{\rm soft}}$ 
(cf.~Eqs.~(\ref{fan}), (\ref{pdf-scr})).

The obtained parton momentum distributions can be used to calculate total and 
diffractive structure functions (SFs) $F_{2}$, $F_{2}^{\rm D(3)}$ 
as\footnote{Strictly speaking, Eq.~(\ref{f2d}) is only appropriate in 
the small $\beta$ limit; at finite $\beta$ the contribution of the
so-called $q\bar q$ diffraction component becomes important.\cite{nik91}} 
\begin{eqnarray}
F_{2}(x,Q^{2}) &=& \sum_{I=q,\bar{q}}e_{I}^{2}\, 
x\, f_{I/p}(x,Q^{2}) \label{f2}\\
F_{2}^{\rm D(3)}(x_{{\mathbb P}},\beta,Q^{2}) &=& 
\sum_{I=q,\bar{q}}e_{I}^{2}\, 
x\, f^{\rm DD}_{I/p}(x,x_{{\mathbb P}},Q^{2}), \label{f2d}
\end{eqnarray}
where $\beta =x/x_{\mathbb P}$ and
$f_{I/p}(x,Q^{2})$, $f^{\rm DD}_{I/p}(x,x_{{\mathbb P}},Q^{2})$ 
are obtained evolving the input distributions  
$f_{I/p}(x,Q_0^{2})$, $f^{\rm DD}_{I/p}(x,x_{{\mathbb P}},Q_0^{2})$
 from $Q^2_0$ to $Q^2$  (for valence quark PDFs 
$q_v(x,Q^2_0)$ a parameterized input (GRV94)\cite{grv94} has been used).

\begin{figure}[ht]
\begin{center}\includegraphics[%
  width=7cm,
  height=4cm]{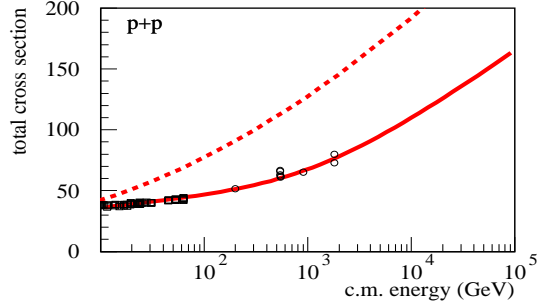}
  \end{center}
\caption{Total $pp$ cross section calculated with and without enhanced diagram
contributions -- full and dashed curves
correspondingly. The compilation of data is from 
Ref.~21.\label{sig-pp}}
\end{figure}

\begin{figure}[ht]
\begin{center}\includegraphics[%
  width=11cm,
  height=7cm]{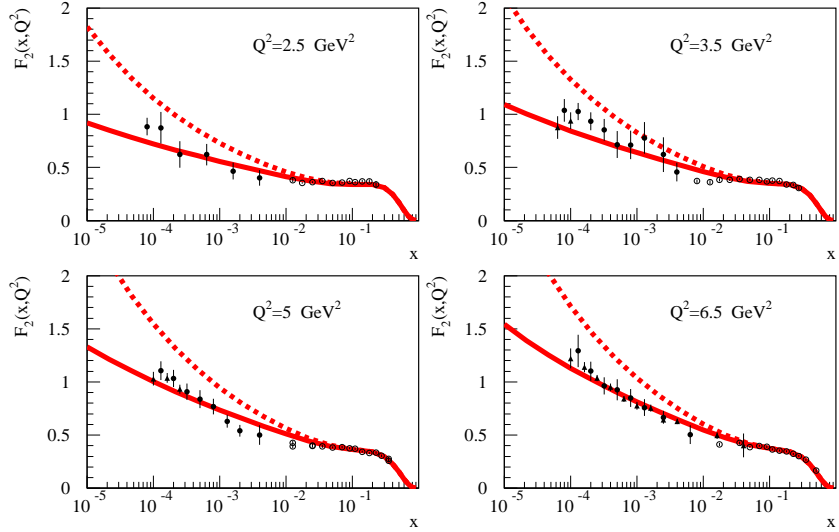}
  \end{center}
\caption{Proton SF $F_{2}(x,Q^{2})$ calculated with and without enhanced 
graph corrections -- full and dashed curves correspondingly. 
The data are from Refs.~22--24.\label{f2comp}}
\end{figure}

\begin{figure}[ht]
\begin{center}\includegraphics[%
  width=11cm,
  height=7cm]{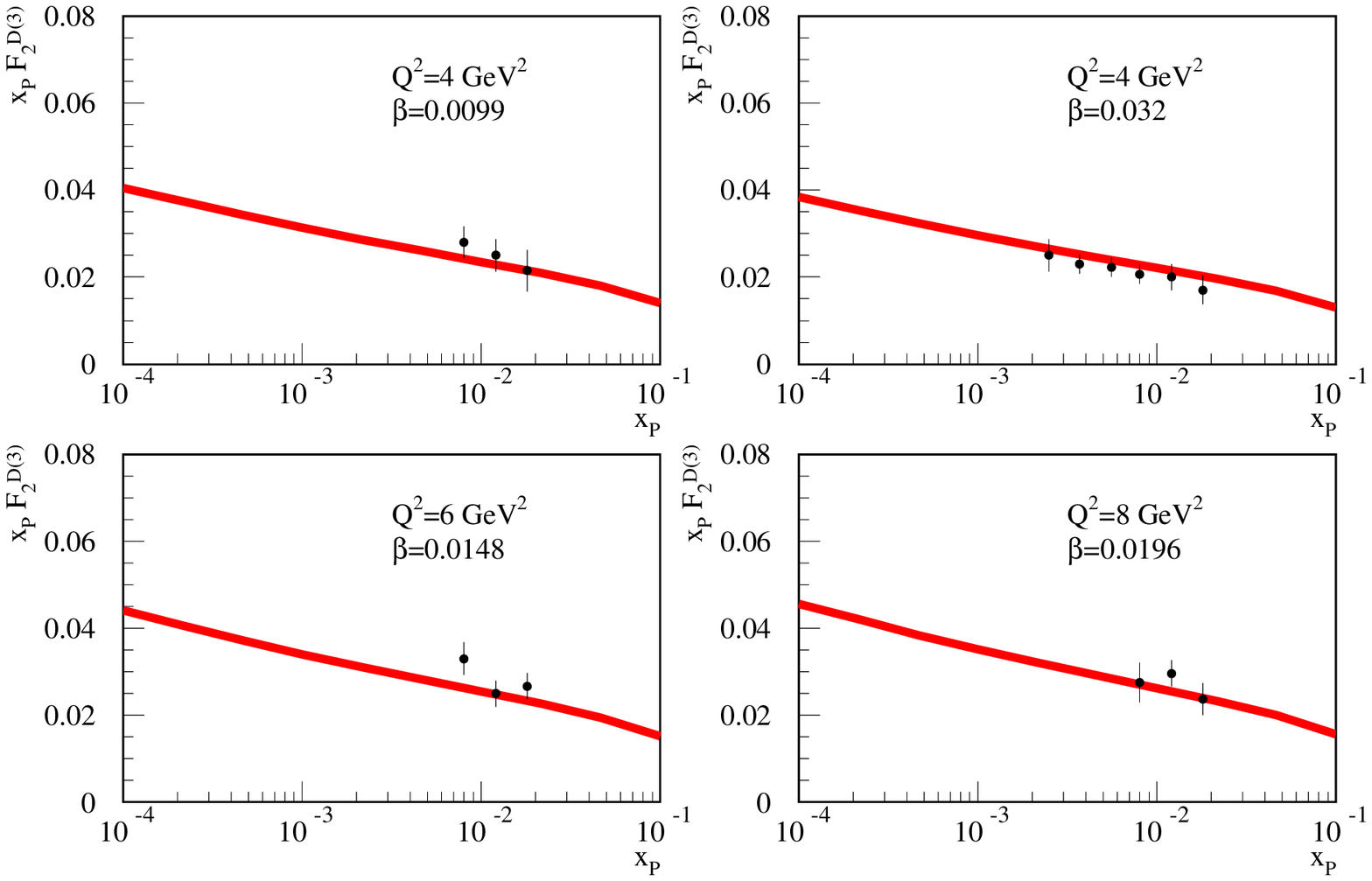}
  \end{center}
\caption{Diffractive proton SF  $F_{2}^{D(3)}(x_{{\mathbb P}},\beta,Q^{2})$ 
in low $\beta$ bins compared to preliminary data of ZEUS 
collaboration (Ref.~25).\label{f3d}}
\end{figure}

For the cutoff value $Q^2_0=2$ GeV$^2$ the Pomeron parameters have been finally 
fixed on the basis of experimental
data on total hadron-hadron cross sections, elastic scattering slopes,
and proton SFs $F_{2}$, $F_{2}^{\rm D(3)}$, in particular, we
had $\alpha_{{\mathbb P}}(0)=0.205$, $\alpha_{{\mathbb P}}'(0)=0.09$ GeV$^{-2}$,
$r_{3{\mathbb P}}=0.026$ GeV$^{-1}$, $\gamma_{\mathbb P}=0.5$ GeV$^{-1}$.
The obtained behavior of $\sigma_{pp}^{{\rm tot}}(s)$, $F_{2}(x,Q^{2})$, and
$F_{2}^{\rm D(3)}(x_{{\mathbb P}},\beta,Q^{2})$ is shown in 
Figs.~\ref{sig-pp}--\ref{f3d}.
In Figs.~\ref{sig-pp}--\ref{f2comp} we plot also 
$\sigma_{pp}^{{\rm tot}}(s)$ and $F_{2}(x,Q^{2})$ as
calculated without enhanced diagram contributions,
 i.e.~setting $r_{3{\mathbb P}}=0$.
 
\begin{figure}[ht]
\begin{center}\includegraphics[%
  width=3cm,
  height=3.3cm]{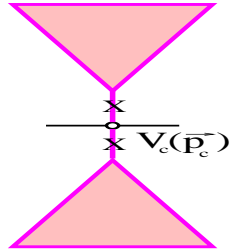}\end{center}
\caption{Diagrams contributing to single inclusive cross sections; 
cut Pomerons are marked by crosses. $V_c(\vec p_c)$ is the particle $c$ 
emission vertex from a cut Pomeron.\label{inclus}}
\end{figure}

It is noteworthy that in the described scheme the contribution of semi-hard 
processes to the interaction eikonal can no longer be expressed in the
usual factorized form, Eqs.~(\ref{G-sh}), (\ref{chi-sh-int}). 
Significant non-factorizable corrections come from graphs where at least 
one Pomeron is exchanged
in parallel to the parton hard process, with the simplest example
given by the 1st diagram on the r.h.s.~in Fig.~\ref{3p-vertex}. In
fact, such contributions are of  importance to get a consistent
description of hadron-hadron cross sections and hadron structure
functions. At the same moment, due to the AGK cancellations\cite{agk} the 
above-mentioned non-factorizable graphs do not
contribute to inclusive high-$p_{t}$ jet spectra.
Single inclusive particle cross sections are defined by
diagrams of Fig.~\ref{inclus}. 
In particular, parton jet spectra are thus obtained in the usual 
factorized form, being defined by the integrand in curly brackets 
in Eq.~(\ref{chi-sh-int}), with the corresponding PDFs 
$f^{\rm tot}_{I/a}(x,Q^{2})$
containing ``fan'' diagram screening corrections.

\section{Results for Extensive Air Showers}

In general, various features of hadronic interactions
 contribute to air shower development in a rather non-trivial
way. Nevertheless, basic EAS observables are grossly defined by a limited 
number of macroscopic characteristics of hadron-air collisions:
inelastic cross sections $\sigma_{h-\rm{air}}^{\rm{inel}}$
, so-called inelasticities $K_{h-\rm{air}}^{\rm{inel}}$
(relative energy differences between the lab.~energies of the
 initial and the most energetic final particles), and charged particle multiplicities
$N_{h-\rm{air}}^{\rm{ch}}$. Indeed, the most fundamental
EAS parameter, the position of the shower maximum $X_{\max}$
(atmospheric depth $X$, where the number of charged particles $N_{e^{\pm}}(X)$
reaches its maximal value),
 is mainly determined by  $\sigma_{h-\rm{air}}^{\rm{inel}}$
 and  $K_{h-\rm{air}}^{\rm{inel}}$. In turn, the number of charged
 particles $N_{e^{\pm}}$, measured at a given observation level, is strongly
 correlated with $X_{\max}$. On the other hand, measured 
 muon number $N_{\mu}$ has a much smaller correlation with the shower maximum
 but depends significantly on  $N_{h-\rm{air}}^{\rm{ch}}$.

\begin{figure}[ht] 
\begin{center}
  \includegraphics[width=7cm,height=4cm,angle=0]{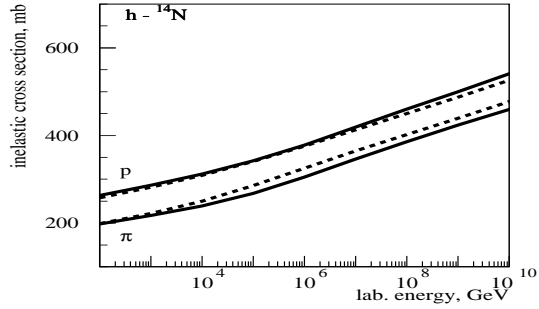} 
\end{center}
   \caption{Inelastic hadron-air cross sections calculated using
 QGSJET-II and QGSJET models
--    full and dashed  curves correspondingly.\label{sigair}} 
\end{figure}

\begin{figure}[ht] 
\begin{center}
  \includegraphics[width=7cm,height=4cm,angle=0]{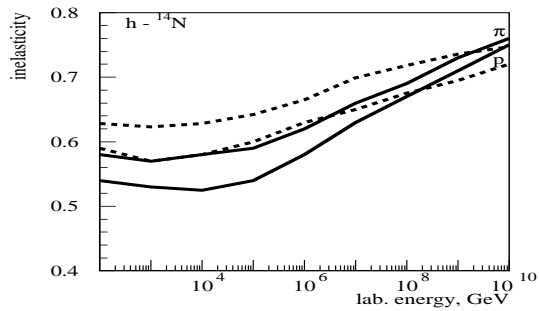} 
\end{center}
   \caption{Inelasticities of  hadron-air interactions  for QGSJET-II and QGSJET
models --     full and dashed  curves correspondingly.\label{kinair}} 
\end{figure}

\begin{figure}[ht] 
\begin{center}
  \includegraphics[width=7cm,height=4cm,angle=0]{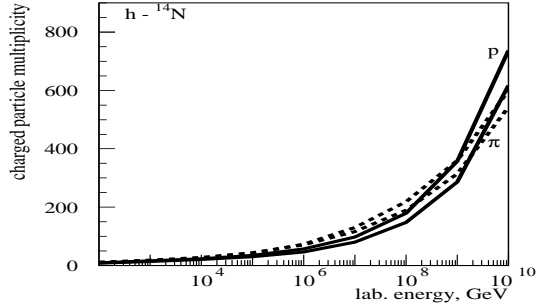} 
\end{center}
   \caption{Multiplicities of charged
particles in  hadron-air interactions for QGSJET-II and QGSJET 
  models - full and dashed curves.\label{mulair}} 
\end{figure}

 In Figs.~\ref{sigair}--\ref{mulair}
  the predictions of the new model for  $\sigma_{h-\rm{air}}^{\rm{inel}}$, 
$K_{h-\rm{air}}^{\rm{inel}}$, and $N_{h-\rm{air}}^{\rm{ch}}$
 are plotted in comparison with the results of the original QGSJET model.
The observed different behavior results from a competition of two effects.
First of all, non-linear  screening corrections reduce the interaction
   eikonal, slow down the energy increase of hadron-air cross section,
compared to the linear scheme with the same Pomeron parameters, and suppress
secondary particle production. On the other hand, larger Pomeron intercept and
 steeper PDFs 
in QGSJET-II\footnote{In the original QGSJET model gluon PDFs are rather flat;
$x\,G(x,Q^2_0)\sim x^{1-\alpha_{{\mathbb P}}(0)}$, with 
$\alpha_{{\mathbb P}}(0)=1.07$, $Q^2_0=4$ GeV$^2$.} 
lead to a faster energy increase of the latter
   quantities. At not too high energies the first effect
   dominates, resulting in  smaller inelasticities and multiplicities.
   Nevertheless, in very high
  energy limit the influence of parton distributions prevails and the new
  model predicts larger values of $K_{h-\rm{air}}^{\rm{inel}}$ and 
$N_{h-\rm{air}}^{\rm{ch}}$.

\begin{figure}[ht] 
\begin{center}
  \includegraphics[width=7cm,height=4cm,angle=0]{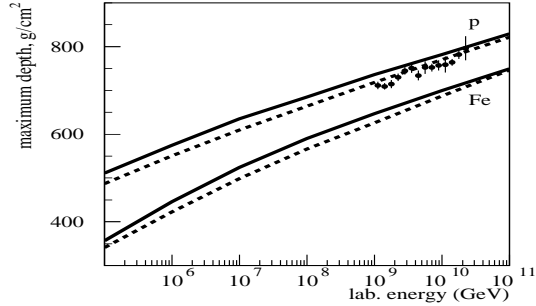} 
\end{center}
   \caption{Average position of the shower maximum for vertical proton-
   and iron-induced EAS as calculated with  QGSJET-II and QGSJET
   models --  full and dashed  curves correspondingly, compared to
    the data  of HIRES collaboration  (points)
     -- Ref.~26.\label{xmaxair}} 
\end{figure}

Sizably smaller inelasticities of the new model lead to a somewhat deeper
position of the shower maximum, compared to the original QGSJET -- 
Fig.~\ref{xmaxair}. As the relative strength of non-linear 
effects is larger for nucleus-nucleus collisions, the difference in $X_{\max}$
 is stronger for nucleus-induced air showers.
Although the effect is rather moderate at highest energies it changes
drastically the interpretation of EAS data: while the predictions of 
the original  QGSJET, being compared to, e.g., HIRES data,\cite{hires} 
 are marginally consistent with the assumption of ultra high energy 
 cosmic rays being only protons, this is no longer the case with the new model.

\begin{figure}[ht] 
\begin{center}
  \includegraphics[width=7cm,height=4cm,angle=0]{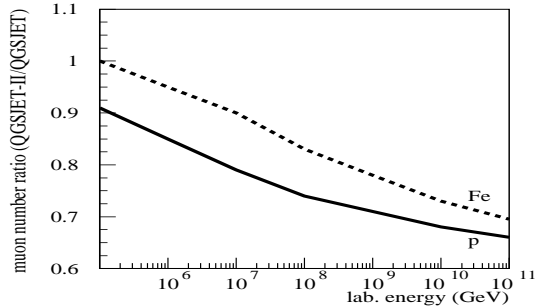} 
\end{center}
   \caption{Relative $N_{\mu}$-difference ($E_{\mu}>1$ GeV) at sea level
    between QGSJET-II and QGSJET models for vertical proton-
   and iron-induced EAS -- full and dashed curves.\label{muair}} 
\end{figure}

The relative difference between QGSJET-II and QGSJET  models
 for predicted muon number 
($E_{\mu}>1$ GeV) at sea level for vertical proton- and iron-induced showers
is shown in Fig.~\ref{muair}. In the new model $N_{\mu}$
 is significantly smaller, by as much as 30\% at highest energies, due to
the substantial reduction of $N_{h-\rm{air}}^{\rm{ch}}$ over a wide energy range
(see Fig.~\ref{mulair}).

In general, applying the new model to EAS data reconstruction
will change present conclusions concerning cosmic ray composition
 towards heavier primaries. At the energies in the region of the ``knee''
  of the CR spectrum ($E_0\sim 10^{15}-10^{16}$ eV) the obtained changes 
  seem to go in the right direction 
  to improve the agreement with experimental data.\cite{hor02,ulr04}
However, at highest energies the existing conflict between the 
composition results obtained with fluorescence light-based measurements 
and with ground arrays\cite{wat04} (much heavier composition in the latter case)
 will aggravate  using the new model. Indeed, the first method
is mainly based on the measured shower maximum position, whereas in the second
case the composition results are obtained from studies of lateral muon 
densities at ground level.
As the predicted reduction of $N_{\mu}$ is much stronger
compared to the corresponding effect for $X_{\max}$, the mismatch between
the corresponding results is expected to increase.

\section{Discussion}

The approach presented here offers a possibility to treat non-linear interaction
effects explicitely in individual high energy hadron-hadron (hadron-nucleus, 
nucleus-nucleus) interactions. The latter are described phenomenologically by
means of enhanced (Pomeron-Pomeron interaction) diagrams. 
A re-summation procedure has been worked
out which allowed to take into account all essential enhanced graphs, both cut
and uncut ones, to obtain positively defined probabilities for various
configurations of inelastic interactions, generally, of complicated topology, 
and to sample such configurations explicitely via a MC method.
As an important feature of the proposed scheme,
the contribution of semi-hard processes to the interaction
eikonal contains an essential non-factorizable part. On the other
hand, by virtue of the AGK cancellations,\cite{agk} the corresponding diagrams
do not contribute to inclusive parton jet spectra. The latter are defined
by the usual QCD factorization ansatz, with screening corrections contained
in parton momentum distributions.

Compared to other approaches,\cite{glr,mcl94} the described scheme does 
not require high parton densities to be reached; essential non-linear 
corrections to the interaction dynamics are consistently taken into account,
 providing a smooth transition between ``dilute'' (small energies, large 
 impact parameters) and ``dense'' regimes and approaching the saturation limit 
 for ``soft'' ($|q^2|<Q^2_0$) particle production in the latter case. 
 On the other hand, the basic assumption of the
scheme -- ``soft'' process dominance of multi-Pomeron vertexes, 
poses restrictions on its application in the region of high parton densities. 
Indeed, after reaching parton density saturation at the chosen cutoff 
scale $Q^2_0$, one has to account 
for corresponding effects in the perturbative ($|q^2|>Q^2_0$) parton dynamics, 
i.e.~to take into account ``hard'' Pomeron-Pomeron coupling. 
Such a treatment is missing here.

\begin{figure}[ht]
\begin{center}\includegraphics[%
  width=7cm,
  height=4cm]{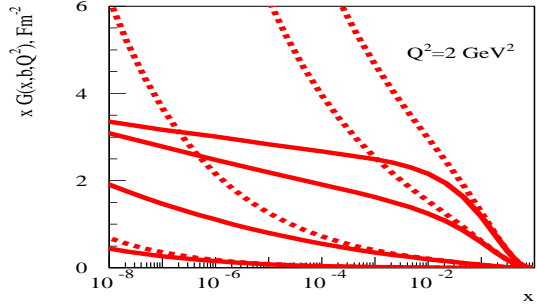}
  \end{center}
\caption{Gluon PDF $x\,G(x,b,Q^2)$ for a given impact parameter $b$
(from up to bottom $b=0,0.5,1,1.5$ Fm) 
calculated with and without enhanced 
graph corrections -- full and dashed curves correspondingly. 
\label{pdf-b}}
\end{figure}

Still, parton saturation is not observed in proton structure functions --
 Fig.~\ref{f2comp}. However, in our scheme the continuing increase of PDFs
in the low $x$ limit is partly due to the non-zero Pomeron slope, which leads
to increasing contribution of proton periphery at small $x$. To investigate
the problem in more detail we plot in Fig.~\ref{pdf-b} gluon momentum
distribution for a given impact parameter $x\,G(x,b',Q^2_0)$, the latter being
defined by the integrand in curly brackets in Eq.~(\ref{pdf-scr}). As seen
from the Figure, at small $b'$ the gluon density is indeed close to the
saturation at the
scale $Q^2_0$ in the small $x$ limit, signalizing the need for perturbative
non-linear corrections.

Applying the approach to air shower simulations we obtained a sizable effect
for the calculated EAS characteristics, with the shower maximum getting
deeper and with the EAS muon content being significantly reduced.
Accounting for perturbative screening corrections (``hard'' Pomeron-Pomeron 
coupling) may enhance corresponding effects. On the other hand, the latter
are only significant in the region of high parton densities, i.e.~in the
``black'' region of small impact parameters. Thus, the likely outcome is
a further reduction of the interaction multiplicity, without a
significant effect on the cross sections and inelasticities, 
correspondingly, on the predicted shower maximum position. 
Moreover, a possible loss of
coherence in the leading hadron system,\cite{dre04} which will only affect
particle production in the fragmentation region, may have an opposite effect:
increasing the inelasticity but having the multiplicity essentially unchanged.
Thus, the two effects will lead to further reduction of EAS muon number while
having $X_{\max}$ essentially unchanged, or even shifted upwards.
Unfortunately, this will worsen the existing contradiction between 
fluorescence light-based and ground-based results on the cosmic ray 
composition.\cite{wat04}
There is a hope to clarify the situation with the data of Pierre Auger 
collaboration,\cite{auger} where both experimental techniques are
available. The latter can provide an indirect
  model consistency check at the highest CR energies.

\section*{Acknowledgments}
The author is grateful to the Organizers for the invitation to this
nice meeting and to R.~Engel and A.~B.~Kaidalov for valuable discussions.
This work has been supported in part  by the German Ministry
for Education and Research (BMBF, Grant 05 CU1VK1/9).

\end{document}